\documentclass[universe,communication,accept,pdftex,moreauthors]{Definitions/mdpi}

\firstpage{1}
\pubvolume{12}
\issuenum{12}
\articlenumber{109}
\pubyear{2026}
\copyrightyear{2026}
\externaleditor{Francesco Terranova}
\datereceived{11 March 2026}
\daterevised{2 April 2026} % Comment out if no revised date
\dateaccepted{8 April 2026} 
\datepublished{9 April 2026}
\usepackage{cases}

\Title{Radiative corrections to the nucleon isovector $g_V$ and $g_A$}

\Author{Oleksandr Tomalak $^{1}$\orcidA{}, Yi-Bo Yang $^{1,2,3,4}$\orcidB{}}

\AuthorNames{Oleksandr Tomalak, Yi-Bo Yang}

\address{%
$^{1}$ \quad Institute of Theoretical Physics, Chinese Academy of Sciences, Beijing 100190, China\\
$^{2}$ \quad University of Chinese Academy of Sciences, School of Physical Sciences, Beijing 100049, China \\
$^{3}$ \quad School of Fundamental Physics and Mathematical Sciences, Hangzhou Institute for Advanced Study, UCAS, Hangzhou 310024, China \\
$^{4}$ \quad International Centre for Theoretical Physics Asia-Pacific, Beijing/Hangzhou 100190, China
}

\corres{Correspondence: tomalak@itp.ac.cn}

\abstract{Electroweak, QCD, and QED radiative corrections to the nucleon low-energy coupling constants $g_V$ and $g_A$ are enhanced by large perturbative logarithms between the electroweak and hadronic scale, as well as between the hadronic scale and the low-energy MeV scale. Additionally, higher-order pion-mass splitting corrections to the nucleon axial-vector charge might be large. By consistently incorporating these effects, we provide an updated relation between the lattice-QCD and physical $g_A$, finding a total radiative correction of $3.5(2.1)\%$ ($5.6(0.7)\%$). This leads to an expected lattice-QCD result of $g^{\mathrm{QCD}}_A = 1.265(26)$ ($g^{\mathrm{QCD}}_A = 1.240(9)$) when based on a combination of lattice-QCD and data-driven (or only data-driven) inputs, respectively. Future phenomenological, chiral perturbation theory, and lattice-QCD studies can improve both the central value and the uncertainty of this estimate.}


\begin{document}

\section{Introduction}

Low-energy charged-current electroweak processes involving nucleons, such as neutron beta decay and inverse beta decay (antineutrino-proton scattering), are characterized by the Fermi coupling constant, the Cabibbo-Kobayashi-Maskawa matrix element $V_{ud}$, and the nucleon isovector vector $g_V$ and axial-vector $g_A$ coupling constants. These processes are now measured with extraordinary precision, with state-of-the-art results provided by the UCN$\tau$~\citep{UCNt:2021pcg}, PERKEO-III~\citep{Markisch:2018ndu,Dubbers:2018kgh}, and JUNO~\citep{JUNO:2015zny,JUNO:2021vlw,JUNO:2022mxj,JUNO:2025gmd} collaborations. For example, the neutron lifetime is known at the $2\times10^{-4}$ level~\citep{UCNt:2021pcg}, and reactor antineutrino experiments such as JUNO are providing increasingly precise data~\citep{JUNO:2015zny,JUNO:2025gmd}.

At this level of experimental accuracy, electroweak, quantum chromodynamics (QCD), and quantum electrodynamics (QED) radiative corrections become essential for extracting fundamental parameters from data~\citep{Cirigliano:2023fnz,Cirigliano:2024nfi,Tomalak:2025jtn,Tomalak:2025okl}. The dominant theoretical uncertainty in these corrections originates from hadronic contributions, whose precise determination remains a central challenge in modern electroweak and hadronic physics~\citep{Marciano:2005ec,Seng:2018qru,Seng:2018yzq,Czarnecki:2019mwq,Hayen:2020cxh,Shiells:2020fqp,Feng:2020zdc,Cirigliano:2023fnz,Ma:2023kfr,Crosas:2025xyv,Cao:2025zxs}. Remarkably, first-principles lattice-QCD simulations are being used now to compute these hadronic effects~\citep{Feng:2020zdc,Ma:2023kfr}, enabling, for instance, a determination of $g_V$ using only Standard Model inputs~\citep{Cirigliano:2023fnz}.

The next major milestone is the first-principles determination of $g_A$ within the Standard Model. Achieving this goal requires precise relations that connect lattice-QCD quantities to their experimentally measured counterparts, a topic of active research~\citep{Cirigliano:2022hob,Cirigliano:2023fnz,Seng:2024ker,Cirigliano:2024nfi}. A proposed relation based on the heavy-baryon chiral perturbation theory (HBChPT)~\citep{Cirigliano:2022hob} suggested that radiative corrections are dominated by leading order (LO) and next-to-leading order (NLO) HBChPT contributions. Subsequent analysis~\citep{Cirigliano:2024nfi} expressed all LO and NLO HBChPT contributions in terms of matrix elements of quark currents, presented  contributions from scales above the energy of hadronic physics, and discussed renormalization group evolution between the scale of hadronic physics and MeV energy. In this work, we improve Ref.~\citep{Cirigliano:2022hob} by including large logarithms and evaluating the unknown low-energy coupling constant. To account for all hadronic corrections, we add contributions from correlation functions involving three currents, as well as several unexplored two‑current correlation functions, to the diagrammatic evaluation of the $\gamma W$ box diagram~\citep{Gorchtein:2021fce}.

In this communication, we numerically analyze large logarithms and hadronic contributions in radiative corrections to $g_A$. We show that the dominant electroweak, QCD, and QED corrections to both $g_V$ and $g_A$ are governed by large perturbative logarithms as well as leading and next-to-leading-order HBChPT contributions from Ref.~\citep{Cirigliano:2022hob}. To guide future Standard Model determinations of $g_A$, we present an updated numerical evaluation of radiative corrections.

%%%%%%%%%%%%%%%%%%%%%%%%%%%%%%%%%%%%%%%%%%
\section{Materials and Methods} \label{sec:methods}

\begin{figure}[H]
\centering
\includegraphics[width=9.0cm]{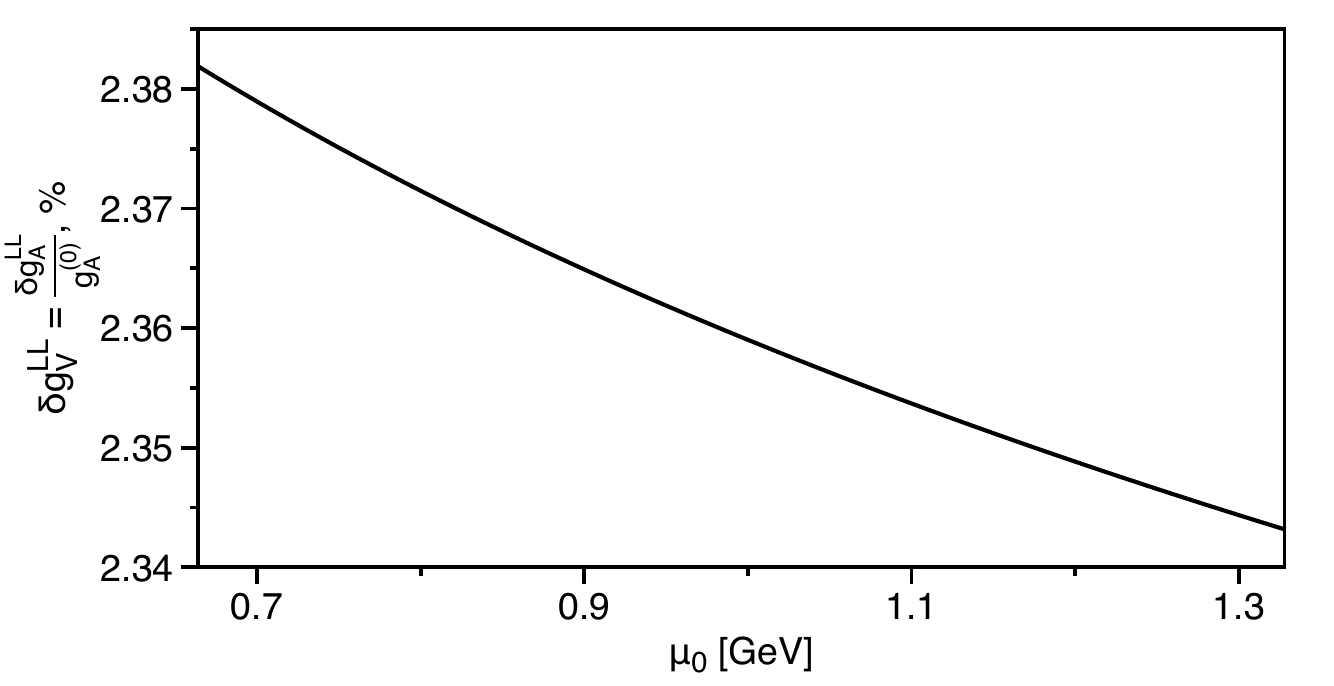}
%} % If the paper is ``preprints'', please uncomment this parenthesis.
\caption{Perturbatively improved relative radiative correction to the nucleon low-energy coupling constants  $\delta g^{\rm LL}_V \left( \mu_\chi = m_e \right)$ and $\frac{\delta g^{\rm LL}_A \left( \mu_\chi = m_e \right)}{g_A^{\left( 0 \right)}}$ is shown as a function of the hadronic scale $\mu_0$.\label{fig:perturbative_improvement_upon_tree_level}}
\end{figure}

We begin our analysis of radiative corrections to $g_V$ and $g_A$ by identifying the leading contributions enhanced by large logarithms. The dominant part of these corrections arises from universal logarithms between the electroweak scale and hadronic scale, as well as between the hadronic scale and the characteristic MeV scale of experiment. In the leading-logarithm (LL) approximation, the corrections at the electron mass scale $\mu_\chi = m_e$ are~\citep{Marciano:2005ec,Cirigliano:2023fnz,Cirigliano:2024nfi}\footnote{We exploit the dimensional regularization in $d = 4 - 2 \varepsilon$ dimensions with the chiral version of modified minimal subtraction scheme subtracting
\begin{equation}
    \frac{1}{\varepsilon} - \gamma_E + \ln \left( 4 \pi \right) + 1.
\end{equation}}
\begin{linenomath}
\begin{equation}
	\delta g^{\mathrm{LL}}_V \left( \mu_\chi = m_e \right) = \frac{\delta g^{\mathrm{LL}}_A \left( \mu_\chi = m_e \right)}{g_A^{\left( 0 \right)}} = \frac{\alpha}{\pi} \left( \ln \frac{M_Z}{\mu_0} - \frac{\alpha_S}{4 \pi} \ln \frac{M_W}{\mu_0} + \frac{3}{4} \ln \frac{\mu_0}{m_e }\right) \approx 2.36(2)\%,\label{eq:gV_leading_logs}
\end{equation}
\end{linenomath}
with the electromagnetic coupling constant $\alpha$ and the strong coupling constant $\alpha_S$. $M_W$ and $M_Z$ denote the electroweak-scale masses of the $W$ and $Z$ bosons, respectively, the hadronic scale is taken as $\mu_0 \approx m_N$, with the nucleon mass $m_N$, and $g_A^{\left( 0 \right)} \approx 1.27$~\citep{Cirigliano:2022hob} denotes the axial-vector charge in the chiral limit without radiative corrections. To estimate the associated uncertainty, we vary $\mu_0$ within the range $\left[ \frac{m_N}{\sqrt{2}}, \sqrt{2} m_N \right]$ and show the dependence of these LL results on the hadronic scale $\mu_0$ in Figure~\ref{fig:perturbative_improvement_upon_tree_level}.

The full one-loop result for the radiative correction becomes independent of the scale $\mu_0$ once the relevant hadronic contributions are taken into account. In the following, we illustrate this explicitly for the vector coupling constant. The one-loop result including the leading QCD logarithm $\delta g^{1-\mathrm{loop}+\mathrm{LL}}_V$ is obtained by accounting for the constant term and hadronic corrections in Eq.~(\ref{eq:gV_leading_logs}) as
\begin{linenomath}
\begin{align}
	\delta g^{1-\mathrm{loop}+\mathrm{LL}}_V \left( \mu_\chi = m_e \right) &= \frac{\alpha}{\pi} \left( \ln \frac{M_Z}{\mu_0} - \frac{\alpha_S}{4 \pi} \ln \frac{M_W}{\mu_0} + \frac{3}{4} \ln \frac{\mu_0}{m_e }\right) - \frac{5 \alpha}{16 \pi} \nonumber \\
	&- e^2 \int \frac{i \mathrm{d}^4 q}{\left( 2 \pi \right)^4} \frac{ \nu^2 + Q^2}{Q^4} \frac{{\overline T}_3 (\nu, Q^2)}{2 m_N \nu},\label{eq:gV_one_loop}
\end{align}
\end{linenomath}
where we take the strong coupling constant $\alpha_S$ at the electroweak matching scale. Here, $Q^2 = - q^2$ denotes the virtuality, and $\nu = v \cdot q = q^0$ is the energy transfer in the nucleon rest frame, with $v=(1,0,0,0)$. The nucleon spin-independent forward Compton scattering invariant amplitude in the isoscalar channel ${\overline T}_3$ is subtracted at large $Q^2$ using a $\mu_0$-dependent operator product expansion (OPE) term~\citep{Cirigliano:2023fnz}:
\begin{linenomath}
\begin{equation}
	{\overline T}_3 (\nu, Q^2) = T_3 (\nu, Q^2) - \frac{4}{3} \frac{m_N \nu}{Q^2 + \mu^2_0} \left( 1 - \frac{\alpha_S}{\pi} \right). \label{eq:T3_OPE}
\end{equation}
\end{linenomath}
The constant term $-\frac{5 \alpha}{16 \pi}$ yields $-0.073\%$, while the hadronic contribution for $\mu_0 = m_N$ is evaluated as $0.156(12)\%$~\citep{Marciano:2005ec,Seng:2018qru,Seng:2018yzq,Czarnecki:2019mwq,Hayen:2020cxh,Shiells:2020fqp,Cirigliano:2023fnz}. The resulting correction to the vector coupling constant,
\begin{linenomath}
\begin{equation}
	\delta g_V ^{1-\mathrm{loop}+\mathrm{LL}} \left( \mu_\chi = m_e \right) = 2.444(12)\%, \label{eq:gV_one_loop_value}
\end{equation}
\end{linenomath}
with an uncertainty estimate from the hadronic input but without perturbative errors, is independent of the hadronic scale $\mu_0$ with a cancellation between logarithms in Eq.~(\ref{eq:gV_one_loop}) and the logarithm from the OPE term in ${\overline T}_3$. The resummation of leading logarithms and inclusion of next-to-leading logarithmic corrections according to Refs.~\citep{Hill:2019xqk,Cirigliano:2023fnz} increases the radiative correction to the vector coupling constant by $0.055\%$. Adding also the uncertainty estimate by varying the matching scales and the hadronic scale $\mu_0$, we obtain
\begin{linenomath}
\begin{equation}
	\delta g_V \left( \mu_\chi = m_e \right) = 2.499(13) \%. \label{eq:gV_value}
\end{equation}
\end{linenomath}

In complete analogy with Eq.~(\ref{eq:gV_one_loop}), the one-loop radiative correction to the axial-vector coupling constant, including the leading QCD logarithm, $\delta g^{1-\mathrm{loop}+\mathrm{LL}}_A$ can be written as
\begin{linenomath}
\begin{align}
	\frac{\delta g^{1-\mathrm{loop}+\mathrm{LL}}_A \left( \mu_\chi = m_e \right)}{g_A^{\left( 0 \right)}} &= \frac{\alpha}{\pi} \left( \ln \frac{M_Z}{\mu_0} - \frac{\alpha_S}{4 \pi} \ln \frac{M_W}{\mu_0} + \frac{3}{4} \ln \frac{\mu_0}{m_e }\right) - \frac{5 \alpha}{16 \pi} + \frac{\delta g_A^{\rm extra}}{g_A^{\left( 0 \right)}} \nonumber \\
	&+ \frac{2 e^2}{g_A^{\left( 0 \right)}} \int \frac{i \mathrm{d}^4 q}{\left( 2 \pi \right)^4} \left( \frac{ \nu^2 - 2 Q^2}{3 Q^2} \frac{\overline{S}_1 (\nu, Q^2)}{Q^2} - \frac{\nu^2}{Q^2 } \frac{{S}_2 (\nu, Q^2)}{m_N \nu} \right), \label{eq:gA_one_loop_AVV_lattice_QCD} \\
	\frac{\delta g_A^{\rm extra}}{g_A^{\left( 0 \right)}} &= \frac{\Delta\overline{g}_A^{\rm QCD + QED} \left(\lambda_\gamma, \mu_0 \right)}{g_A^{\left( 0 \right)}} + e^2 \int \frac{i \mathrm{d}^4 q}{(2\pi)^4} \frac{2 \overline{m}{t}_{VP} \left(q, v \right) - {\overline t}_{VA} \left(q, v \right)}{ q^2 - \lambda_\gamma^2}. \label{eq:gA_extra_one_loop_AVV_lattice_QCD}
\end{align}
\end{linenomath}
The OPE-subtracted nucleon spin-dependent forward Compton scattering invariant amplitudes in the isovector channel ${\overline S}_1$ and ${S}_2$ yield a correction of $0.169(6)\%$~\citep{Gorchtein:2021fce} for $\mu_0 = m_N$. The term $\delta g_A^{\rm extra}$, often omitted in precise evaluations~\citep{Gorchtein:2021fce}, must be included for a complete result. Here, $\Delta\overline{g}_A^{\rm QCD + QED} \left(\lambda_\gamma, \mu_0 \right)$ is the OPE-subtracted QED radiative correction to the nucleon matrix element of the axial-vector quark current, evaluated in the Feynman gauge, $\xi = 1$,\footnote{An extra term has to be added to relate $\Delta\overline{g}_A^{\rm QCD + QED} \left(\lambda_\gamma, \mu_0 \right)$ to the OPE-subtracted QED radiative correction in an arbitrary $R_\xi$ gauge $\Delta\overline{g}_A^{\rm QCD + QED} \left(\xi, \lambda_\gamma, \mu_0 \right)$,
\begin{linenomath}
\begin{equation}
	\frac{\Delta\overline{g}_A^{\rm QCD + QED} \left(\lambda_\gamma, \mu_0 \right)}{g_A^{\left( 0 \right)}} = \frac{\Delta\overline{g}_A^{\rm QCD + QED} \left(\xi, \lambda_\gamma, \mu_0 \right)}{g_A^{\left( 0 \right)}}+ \frac{1 - \xi}{2} \frac{\alpha}{\pi} \left[ \ln \frac{\lambda^2_\gamma}{\mu^2_0} +\frac{ a \xi \ln \left( a \xi \right)}{1- a \xi} \right], \label{eq:gA_QED_arbitrary_gauge}
\end{equation}
\end{linenomath}
with the arbitrary parameter $a$ that enters the gauge-dependent piece of the photon propagator as $1/ \left( q^2 - a \xi \lambda_\gamma^2 \right)$.} with infrared regularization by the photon mass $\lambda_\gamma$. The integral of the second term in $\delta g_A^{\rm extra}$ involves OPE-subtracted two-current correlation functions: vector-pseudoscalar ${t}_{VP}$, with the isospin-averaged quark mass $\overline{m}$, and vector-axial-vector ${\overline t}_{VA}$, that cancel both the infrared divergence and residual $\mu_0$ dependence from $\Delta\overline{g}_A^{\rm QCD + QED} \left(\lambda_\gamma, \mu_0 \right)$. The explicit definitions for hadronic objects ${t}_{VP}$ and ${\overline t}_{VA}$ are presented in Ref.~\citep{Cirigliano:2024nfi}. In this work, we provide the first numerical estimates of the term $\delta g_A^{\rm extra}$.

The term $\delta g_A^{\rm extra}$ can be evaluated solely in lattice QCD. However, it is instructive to discuss dominant contributions to this object. First, the leading contributions in $\delta g_A^{\rm extra}$ from one-pion intermediate states together with a corresponding dependence on the nucleon isovector magnetic moment, cancel exactly between two terms in $\delta g_A^{\rm extra}$~\citep{Cirigliano:2024nfi}. Two-pion intermediate states generate a large logarithm as well as the next-to-leading order contribution, expressed in terms of the NLO HBChPT low-energy coupling constants (LECs) \(c_3\) and \(c_4\), as
\begin{linenomath}
\begin{align}
	\frac{\delta g_A^{\rm extra}}{g_A^{\left( 0 \right)}} & \approx \frac{Z_\pi \left[ 1 + 3 \left( g_A^{\left( 0 \right)} \right)^2 \right] }{2} \frac{\alpha}{\pi} \ln \frac{m_N}{m_\pi} + 2 \alpha Z_\pi m_\pi \left[ c_4 - c_3 + \frac{3}{8 m_N} + \frac{9}{16} \frac{\left( g_A^{\left( 0 \right)} \right)^2}{m_N} \right] \nonumber \\
	&+ \frac{\delta g_A^{\rm extra, N}}{g_A^{\left( 0 \right)}} + \frac{\delta g_A^{\rm extra, inel}}{g_A^{\left( 0 \right)}} \nonumber \\
    &\approx
		\begin{cases}
			2.9(0.3)\%, ~~~c_{4}^{\rm N^3LO} - c_{3}^{\rm N^3LO} \\
			2.2(0.2)\%, ~~~c_{4}^{\rm NLO} - c_{3}^{\rm NLO} \\
			1.6(0.1)\%, ~~~ c_{4}^{\rm Born} - c_{3}^{\rm Born} \\
            		0.7(2.1)\%, \left( 2c_4-c_3 \right)^{\rm LQCD} - c_4^{\rm N^3LO}
		\end{cases}
		\hspace{-0.45cm} + \frac{\delta g_A^{\rm extra, N}}{g_A^{\left( 0 \right)}} + \frac{\delta g_A^{\rm extra, inel}}{g_A^{\left( 0 \right)}}, \label{eq:gA_extra_one_loop_AVV_lattice_QCD_pion_and_two_pion_corrections}
\end{align}
\end{linenomath}
where we have chosen the nucleon mass as the ultraviolet cutoff in the logarithmic term, and $m_\pi$ denotes the pion mass. The uncertainty is estimated by adding, in quadrature, the power-counting uncertainty of the NLO contribution to the uncertainty on the coupling constant $c_4 - c_3$. The parameter $Z_\pi \approx 0.81$ is the QED isospin-breaking LEC responsible for the electromagnetic pion-mass splitting.

The NLO HBChPT LECs \(c_3\) and \(c_4\) enter the second term. Phenomenological determinations from \(\pi N\) scattering data~\citep{Hoferichter:2015tha,Hoferichter:2015hva} provide the following values (errors added in quadrature):
\begin{linenomath}
\begin{align}
	c_3^{\rm NLO} &= -3.61(5)~\mathrm{GeV}^{-1}, ~ c_4^{\rm NLO} = 2.17(3)~\mathrm{GeV}^{-1}, ~~~~ c_4^{\rm NLO}-c_3^{\rm NLO} = 5.78(6)~\mathrm{GeV}^{-1}, \\
	c_3^{\rm N^3LO} &= -5.61(6)~\mathrm{GeV}^{-1},  c_4^{\rm N^3LO} = 4.26(4)~\mathrm{GeV}^{-1}, ~ c_4^{\rm N^3LO}-c_3^{\rm N^3LO} = 9.87(7)~\mathrm{GeV}^{-1}.
\end{align}
\end{linenomath}
The fit results change significantly going from NLO to N\(^2\)LO but stabilize at N\(^3\)LO~\citep{Hoferichter:2015tha,Hoferichter:2015hva}. Since our calculation is performed at NLO in HBChPT, we estimate the systematic uncertainty from the difference between the NLO and the more complete N\(^3\)LO values (rather than using N\(^2\)LO). The combination \(2 c_4^{\rm NLO} - c_3^{\rm NLO} = 7.95(8)~\mathrm{GeV}^{-1}\) (\(2 c_4^{\rm N^3LO} - c_3^{\rm N^3LO} = 14.13(0.10)~\mathrm{GeV}^{-1}\)) can also be extracted from the \(m_\pi^3\) dependence of the nucleon axial-vector charge \(g_A\). However, a recent lattice-QCD calculation~\citep{Hall:2025ytt} gives a value an order of magnitude smaller \( \left( 2c_4-c_3 \right)^{\rm LQCD} = 0.85(0.25)~\mathrm{GeV}^{-1}\). Using this lattice-QCD result together with the phenomenological extraction of \(c_3\) or \(c_4\) as an input, we obtain the relevant combination of LECs \( c_4 - c_3 = \left( 2c_4-c_3 \right)^{\rm LQCD} - c_4^{\rm NLO}  = -1.32(0.25)~\mathrm{GeV}^{-1}\) and \( c_4 - c_3 = \left( 2c_4-c_3 \right)^{\rm LQCD} - c_4^{\rm N^3LO} = -3.41(0.25)~\mathrm{GeV}^{-1}\). For the central value of the QED correction with lattice-QCD input $\left( 2c_4-c_3 \right)^{\rm LQCD}$, we adopt the LEC \(c_4^{\rm N^3LO}\): \(c_4 - c_3 = \left( 2c_4-c_3 \right)^{\rm LQCD} - c_4^{\rm N^3LO}  = -3.4(13.3)~\mathrm{GeV}^{-1}\), and assign a systematic uncertainty by taking the difference with the value obtained from the largest alternative combination resulting in the largest error estimate in our evaluations. We also apply, for the first time, the Born approximation with only the nucleon intermediate state to determine the NLO HBChPT LECs from the expressions in Appendix D of Ref.~\citep{Cirigliano:2024nfi}. We then evaluate the LECs of interest as\footnote{The other NLO HBChPT LECs are $c^{\rm Born}_2 = - \frac{3}{8} \frac{\left( g_A^{\left( 0 \right)}\right)^2}{m_N}+ \frac{\pi m_\pi}{\left( 4 \pi F_\pi \right)^2} = -0.3(0.1)$ and $\kappa^{\rm Born}_1 = m_N \frac{\pi m_\pi}{\left( 4 \pi F_\pi \right)^2} = 1.9(0.3)$. Exploiting the constraint on the nucleon isovector magnetic moment $\kappa_1$, the Wilson coefficient $c_4$ can be estimated as $c_4 = \frac{\kappa_1}{4 m_N} -\frac{\left( g_A^{\left( 0 \right)}\right)^2}{4 m_N} + \left( g_A^{\left( 0 \right)}\right)^4 \frac{\pi m_\pi}{\left( 4 \pi F_\pi \right)^2} = 1.4(0.4)~\mathrm{GeV}^{-1}$ validating our uncertainty estimate.}
\begin{linenomath}
\begin{align}
	c^{\rm Born}_3 &= \frac{\left( g_A^{\left( 0 \right)}\right)^2}{4 m_N} - \left[ \frac{3}{2} + \left( g_A^{\left( 0 \right)}\right)^2 \right]  \left( g_A^{\left( 0 \right)}\right)^2 \frac{\pi m_\pi}{\left( 4 \pi F_\pi \right)^2} = -1.2(0.4)~\mathrm{GeV}^{-1}, \\
	c^{\rm Born}_4 &= -\frac{\left( g_A^{\left( 0 \right)}\right)^2}{4 m_N} +  \left[ 1+ \left( g_A^{\left( 0 \right)}\right)^2 \right]  \left( g_A^{\left( 0 \right)}\right)^2 \frac{\pi m_\pi}{\left( 4 \pi F_\pi \right)^2} = 0.9(0.2)~\mathrm{GeV}^{-1},\label{eq:c3c4_Born}
\end{align}
\end{linenomath}
with the pion decay constant $F_\pi \approx 92.4~\mathrm{MeV}$, and estimate the error as the nucleon state contribution to these LECs. The Born result for $2 c^{\rm Born}_4 - c^{\rm Born}_3 = 3.0(0.8)~\mathrm{GeV}^{-1}$ is also in tension both with lattice QCD~\citep{Hall:2025ytt} and fits to the experimental data~\citep{Hoferichter:2015tha,Hoferichter:2015hva}. Thus, it is worthwhile to explore whether additional combinations of \(c_3\) and \(c_4\) can be constrained using the light-quark mass dependence of isoscalar nucleon charges, such as isoscalar scalar charge, as well as to determine \(c_3\) and \(c_4\) directly from nucleon matrix elements as described in Appendix D of Ref.~\citep{Cirigliano:2024nfi}. Isovector charges have been precisely determined from lattice QCD in Ref.~\citep{Wang:2025nsd}. Moreover, a lattice-QCD determination of $\delta g_A^{\rm extra}$ would directly constrain the low-energy coupling constant $c_4-c_3$ and clarify the HBChPT convergence pattern.

For estimating the size of hadronic corrections, we separate the Born contribution from nucleon states, $\delta g_A^{\rm extra, N}$, in Eq.~(\ref{eq:gA_extra_one_loop_AVV_lattice_QCD_pion_and_two_pion_corrections}), and indicate inelastic contributions as $\delta g_A^{\rm extra, inel}$. According to evaluations of typical hadronic objects in Refs.~\citep{Tomalak:2017shs,Tomalak:2018jak,Tomalak:2018uhr,Tomalak:2018dho,Cirigliano:2023fnz}, the nucleon intermediate state contribution is either dominant or has a comparable size to inelastic excitations. We get $-0.0127(2)\%$ and $0.222(3)\%$ for the nucleon-state contribution to two terms in $\delta g_A^{\rm extra}$ of Eq.~(\ref{eq:gA_extra_one_loop_AVV_lattice_QCD}), respectively. We describe the details of this calculation in Appendix~\ref{app:nucleon_state} and investigate Ward identities for the nucleon state by evaluating hadronic objects from an alternative decomposition of radiative corrections in Appendix~\ref{app:Ward_identity}.

Accounting for the constant term and the hadronic contribution from the second line of Eq.~(\ref{eq:gA_one_loop_AVV_lattice_QCD}), we obtain the one-loop radiative correction, including the leading QCD logarithm, to the axial-vector coupling constant $\delta g^{1-\mathrm{loop}+\mathrm{LL}}_A$,
\begin{linenomath}
\begin{equation}
	\frac{\delta g^{1-\mathrm{loop}+\mathrm{LL}}_A \left( \mu_\chi = m_e \right)}{g_A^{\left( 0 \right)}} = 2.457(6)\% + \frac{\delta g_A^{\rm extra}}{g_A^{\left( 0 \right)}} = 
		\begin{cases}
			5.5(0.7)\%, \quad c_{4}^{\rm N^3LO} - c_{3}^{\rm N^3LO} \\
			4.9(0.7)\%, \quad c_{4}^{\rm NLO} - c_{3}^{\rm NLO} \\
			4.3(0.4)\%, \quad c_{4}^{\rm Born} - c_{3}^{\rm Born} \\
            		3.4(2.1)\%, ~\left( 2c_4-c_3 \right)^{\rm LQCD} - c_4^{\rm N^3LO}
		\end{cases}. \label{eq:gA_one_loop_AVV_lattice_QCD_number}
\end{equation}
\end{linenomath}
For the uncertainty estimate in the first two lines (the third line), we add $0.2\%$ uncertainty from the variation of the hadronic scale under the logarithm in Eq.~(\ref{eq:gA_extra_one_loop_AVV_lattice_QCD_pion_and_two_pion_corrections}) to $0.7\%$ difference between results with NLO and N\(^3\)LO couplings ($0.1\%$ from the nucleon state contribution in $c_{4}^{\rm Born} - c_{3}^{\rm Born}$) and to $0.3\%$ difference in the evaluation of the nucleon intermediate state contribution between Appendix~\ref{app:nucleon_state} and Appendix~\ref{app:Ward_identity} in quadrature, respectively. Compared to previous estimates~\citep{Gorchtein:2021fce}, this analysis consistently incorporates the large logarithmic enhancements at the one-loop level, explicitly accounts for contributions from all two- and three-current correlation functions, and includes the associated uncertainty.

As in the case of the vector coupling constant, we further include the next-to-leading logarithmic corrections and resummation, and obtain\footnote{For reference, we also present the correction $\delta g_A$ at the chiral scale of the nucleon mass
\begin{linenomath}
\begin{equation}
	\frac{\delta g_A \left( \mu_\chi = m_N \right)}{g_A^{\left( 0 \right)}} = 
		\begin{cases}
			4.2(0.7)\%, \quad c_{4}^{\rm N^3LO} - c_{3}^{\rm N^3LO} \\
			3.6(0.7)\%, \quad c_{4}^{\rm NLO} - c_{3}^{\rm NLO} \\
			3.0(0.4)\%, \quad c_{4}^{\rm Born} - c_{3}^{\rm Born} \\
            		2.1(2.1)\%, ~\left( 2c_4-c_3 \right)^{\rm LQCD} - c_4^{\rm N^3LO}
		\end{cases}.
\end{equation}
\end{linenomath}}
\begin{linenomath}
\begin{equation}
	\frac{\delta g_A \left( \mu_\chi = m_e \right)}{g_A^{\left( 0 \right)}} = 2.513(8)\% + \frac{\delta g_A^{\rm extra}}{g_A^{\left( 0 \right)}} = 
		\begin{cases}
			5.6(0.7)\%, \quad c_{4}^{\rm N^3LO} - c_{3}^{\rm N^3LO} \\
			5.0(0.7)\%, \quad c_{4}^{\rm NLO} - c_{3}^{\rm NLO} \\
			4.4(0.4)\%, \quad c_{4}^{\rm Born} - c_{3}^{\rm Born} \\
            		3.5(2.1)\%, ~\left( 2c_4-c_3 \right)^{\rm LQCD} - c_4^{\rm N^3LO}
		\end{cases}. \label{eq:gA0_AVV_lattice_QCD_number}
\end{equation}
\end{linenomath}

%%%%%%%%%%%%%%%%%%%%%%%%%%%%%%%%%%%%%%%%%%
\section{Results and Discussion} \label{sec:discussion}

It is instructive to separate the contributions from the uncertain $c_4 - c_3$, the remaining uncertain hadronic corrections, and the precisely calculable perturbative contribution. The resulting radiative correction can be decomposed into three terms:
\begin{linenomath}
\begin{equation}
	\frac{g_A}{g_A^{\mathrm{QCD}} g_V} -1 = -0.033(15)\% \;+\; 2 \alpha Z_\pi m_\pi (c_4 - c_3) \;+\; \frac{\delta g_A^{\rm had}}{g_A^{\mathrm{QCD}} g_V}, \label{eq:result}
\end{equation}
\end{linenomath}
where $g_A^{\mathrm{QCD}}$ denotes the axial-vector charge in the isospin limit. The first term contains the difference of $\gamma W$ box contributions to $g_A$ and $g_V$, as well as higher-order remnants.\footnote{Explicitly, these remnants are $\frac{\delta g_A - \delta g_A^{\rm extra}}{g_A^{\left( 0 \right)} g_V} \left( \frac{g_A^{\left( 0 \right)}}{g_A^{\mathrm{QCD}}}- 1 \right) = -0.110\%$, where we take the HBChPT input from Refs.~\citep{Kambor:1998pi,Bernard:2006te}, $\left| \frac{\delta g_A - \delta g_A^{\rm extra} - \left( g_V - 1 \right)}{g_A^{\left( 0 \right)}} \left(\frac{1}{g_V} - 1 \right) \right|  < 0.0004\%$, and $\frac{1 - C^r_\beta}{g_V} \left( \left( C^r_\beta - \delta g_A \right)\frac{g_A^{\left( 0 \right)}}{g_A^{\mathrm{QCD}}} - 1 \right) = \left( 0.058-0.075 \right)\%$ for a range of values $\delta g_A^{\rm extra}$ in this work. $C^r_\beta = 1 + \mathcal{O} \left( \alpha \right)$ is the semileptonic Wilson coefficient in the low-energy effective field theory of quarks and leptons, evaluated at the $\mathrm{GeV}$ renormalization scale.} The second term depends on the combination \(c_4 - c_3\) of the NLO HBChPT LECs \(c_3\) and \(c_4\). This term has a sizable uncertainty as discussed in the previous section, with the coefficient \(2 \alpha Z_\pi m_\pi = 0.16(2)~\mathrm{GeV}\) and an uncertainty estimate from the HBChPT power counting. The third term represents the hadronic contributions $\delta g_A^{\rm extra}$ but explicitly removes the dependence on \(c_4 - c_3\),
\begin{linenomath}
\begin{equation}
	\frac{\delta g_A^{\rm had}}{g_A^{\mathrm{QCD}} g_V} = \frac{\delta g_A^{\rm extra}}{g_A^{\mathrm{QCD}} g_V} - 2 \alpha Z_\pi m_\pi \left( c_4 - c_3 \right) \approx 1.5(0.4)\%.
\end{equation}
\end{linenomath}
The corresponding uncertainty is given by $0.3\%$ difference in the evaluation of the nucleon intermediate state contribution between Appendix~\ref{app:nucleon_state} and Appendix~\ref{app:Ward_identity} and $0.2\%$ uncertainty from the variation of the hadronic scale under the logarithm in Eq.~(\ref{eq:gA_extra_one_loop_AVV_lattice_QCD_pion_and_two_pion_corrections}) added in quadrature. The correction in Eq.~(\ref{eq:result}) does not depend on the scale and is free from large logarithms of Eq.~(\ref{eq:gV_leading_logs}).

For various $c_4 - c_3$ inputs considered above, we obtain
\begin{linenomath}
\begin{equation}
	\frac{g_A}{g_A^{\mathrm{QCD}} g_V} -1 = 
		\begin{cases}
			3.0(0.7)\%, \quad c_{4}^{\rm N^3LO} - c_{3}^{\rm N^3LO} \\
			2.3(0.7)\%, \quad c_{4}^{\rm NLO} - c_{3}^{\rm NLO} \\
			1.8(0.4)\%, \quad c_{4}^{\rm Born} - c_{3}^{\rm Born} \\
            		0.8(2.1)\%, ~\left( 2c_4-c_3 \right)^{\rm LQCD} - c_4^{\rm N^3LO}
		\end{cases}. \label{eq:gA_AVV_lattice_QCD_number}
\end{equation}
\end{linenomath}
As our final result for relating lattice-QCD determinations to experimental measurements, we quote Eq.~(\ref{eq:gA_AVV_lattice_QCD_number}). Based on the current experimental value of the axial-vector to vector coupling constant ratio $\frac{g_A}{g_V} = 1.2753(13)$~\citep{ParticleDataGroup:2024cfk} and N\(^3\)LO LECs~\citep{Hoferichter:2015tha,Hoferichter:2015hva}, we expect the lattice-QCD result for the axial-vector charge $g^{\mathrm{QCD}}_A = 1.240(9)$. Exploiting the strong lattice-QCD constraint on NLO HBChPT LECs and estimating the systematic uncertainty by taking the largest difference between alternative inputs, we get $g^{\mathrm{QCD}}_A = 1.265(26)$.

\begin{figure}[H]
\centering
\includegraphics[width=9.0cm]{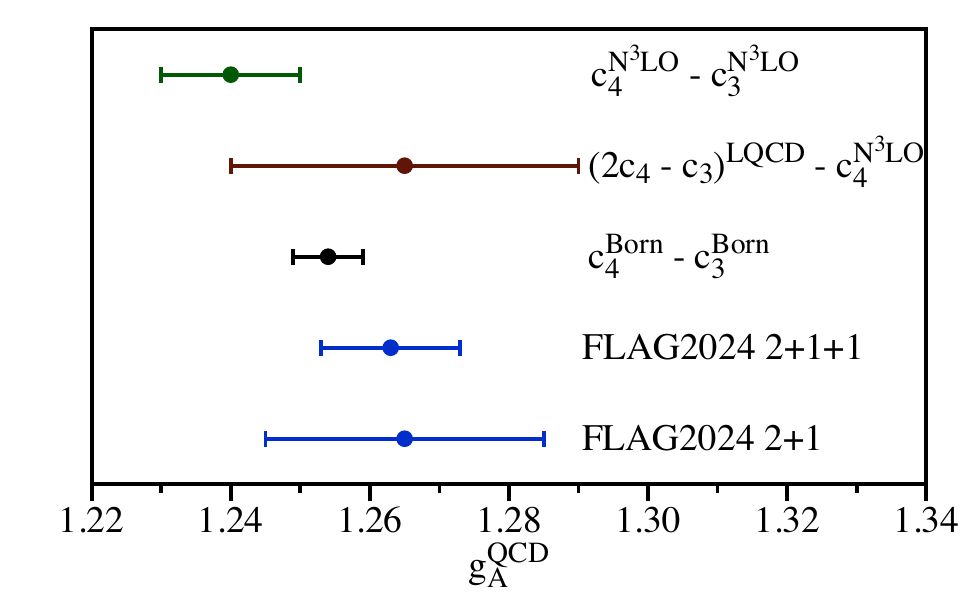}
%} % If the paper is ``preprints'', please uncomment this parenthesis.
\caption{The expected lattice-QCD result for the nucleon axial-vector charge, based on the experimental measurements of $g_A$~\citep{ParticleDataGroup:2024cfk} and up-to-date radiative corrections evaluated in this work, is compared to the Flavour Lattice Averaging Group (FLAG) average of lattice-QCD calculations~\citep{FlavourLatticeAveragingGroupFLAG:2024oxs,Chang:2018uxx,Walker-Loud:2019cif,Jang:2023zts,Alexandrou:2023qbg,Liang:2018pis,Park:2021ypf,QCDSFUKQCDCSSM:2023qlx,Bali:2023sdi,Djukanovic:2024krw,Harris:2019bih,RQCD:2019jai,Djukanovic:2022wru,Gupta:2018qil,Bhattacharya:2015wna}. The results of this work are presented for three cases: using NLO HBChPT coupling constants in radiative corrections from N$^3$LO fits to $\pi N$ scattering data; incorporating the available lattice-QCD constraint from Ref.~\citep{Hall:2025ytt}; and using estimates from the Born approximation.\label{fig:gA_QCD}}
\end{figure}

We present our results for the nucleon axial-vector charge in Fig.~\ref{fig:gA_QCD} and compare them to the Flavour Lattice Averaging Group (FLAG) average of lattice-QCD calculations~\citep{FlavourLatticeAveragingGroupFLAG:2024oxs,Chang:2018uxx,Walker-Loud:2019cif,Jang:2023zts,Alexandrou:2023qbg,Liang:2018pis,Park:2021ypf,QCDSFUKQCDCSSM:2023qlx,Bali:2023sdi,Djukanovic:2024krw,Harris:2019bih,RQCD:2019jai,Djukanovic:2022wru,Gupta:2018qil,Bhattacharya:2015wna}, which itself comprises data from both 2+1+1 and 2+1 flavour simulations distinguishing between calculations with and without a dynamical charm quark in the sea. We find a tension between our result based on N$^3$LO fits for the coupling constants $c_3$ and $c_4$ and modern lattice-QCD determinations. Including the lattice-QCD constraint on $2c_4 - c_3$~\citep{Hall:2025ytt} and inflating the error bars or estimating the LECs $c_3$ and $c_4$ within the Born approximation removes the tension. This situation underscores the need for a direct lattice-QCD determination of the hadronic QED corrections to $g_A$ and/or for improved lattice-QCD constraints on the next-to-leading-order HBChPT low-energy coupling constants $c_{3}$ and $c_4$, as well as for improvements in phenomenological and ChPT analyses.

%%%%%%%%%%%%%%%%%%%%%%%%%%%%%%%%%%%%%%%%%%
%\authorcontributions{Conceptualization, O.T.; methodology, O.T.; software, O.T.; validation, O.T. and Y.Y.; formal analysis, O.T.; investigation, O.T.; resources, O.T.; data curation, O.T.; writing---original draft preparation, O.T.; writing---review and editing, O.T. and Y.Y.; visualization, O.T.; supervision, O.T.; project administration, O.T. and Y.Y.; funding acquisition, O.T. and Y.Y.. All authors have read and agreed to the published version of the manuscript.}

\authorcontributions{All authors have read and agreed to the published version of the manuscript.}

\funding{This research was funded by the National Science Foundation of China under Grants No. 12347105, 12525504, 12293060, 12293062, and 12447101.}

\dataavailability{No new data were created.}

\acknowledgments{Wolfram Mathematica 12.0~\citep{Wolfram:2022Mathematica} and DataGraph 5.5~\citep{MacAskill:2012JSS} were extremely useful in this work.}

\conflictsofinterest{The authors declare no conflicts of interest.}

\newpage
\abbreviations{Abbreviations}{
The following abbreviations are used in this manuscript:
\\

\noindent 
\begin{tabular}{@{}ll}
FLAG & Flavour Lattice Averaging Group\\
HBChPT & Heavy Baryon Chiral Perturbation Theory\\
LEC & Low-Energy Coupling constant\\
LL & Leading Logarithm\\
LO & Leading Order\\
LQCD & Lattice Quantum ChromoDynamics\\
NLO & Next-to-Leading Order\\
MeV & MegaelectronVolt\\
OPE & Operator Product Expansion\\
QCD & Quantum ChromoDynamics\\
QED & Quantum ElectroDynamics
\end{tabular}
}

%%%%%%%%%%%%%%%%%%%%%%%%%%%%%%%%%%%%%%%%%%
%% Optional
\appendixtitles{yes} % Leave argument "no" if all appendix headings stay EMPTY (then no dot is printed after "Appendix A"). If the appendix sections contain a heading then change the argument to "yes".
\appendixstart
\appendix
\section[\appendixname~\thesection]{Nucleon-state contribution to $\delta g_A^{\rm extra}$} \label{app:nucleon_state}

In this Appendix, we present the evaluation of the nucleon-state contribution $\delta g_A^{\rm extra, N}$ to $\delta g_A^{\rm extra}$ in Eq.~(\ref{eq:gA_extra_one_loop_AVV_lattice_QCD}):
\begin{linenomath}
\begin{align}
	&\frac{\delta g_A^{\rm extra}}{g_A^{\left( 0 \right)}} = \left[ \frac{e^2 \overline{m} b_\rho b_\lambda}{g_A^{\left( 0 \right)}} \int \frac{i \mathrm{d}^4 q}{(2\pi)^4} \frac{g_{\mu \nu} \left( \overline{t}^{\mu \nu \lambda \rho}_{AVV} \left( q, v \right) + \overline{t}^{\mu \nu \lambda \rho}_{AVV,0}\left( q, v \right) \right)}{q^2 - \lambda_\gamma^2} + 2 i \pi \delta \left( v \cdot r \right) \Delta_\mathrm{em} m_N \right]\Bigg |_{r_\lambda = 0} \nonumber \\
    &+ e^2 \int \frac{i \mathrm{d}^4 q}{(2\pi)^4} \frac{2 \overline{m}{t}_{VP} \left(q, v \right) - {\overline t}_{VA} \left(q, v \right)}{ q^2 - \lambda_\gamma^2},\label{eq:gA_extra_one_loop_AVV}
\end{align}
\end{linenomath}
with the momentum insertion in the axial-vector current $r_\lambda$, an arbitrary vector $b$: $b^2 = -1,~b \cdot v = 0$, and the nucleon electromagnetic correction to the isospin-averaged nucleon mass,
\begin{linenomath}
\begin{equation}
	\Delta_\mathrm{em} m_N = -\frac{e^2}{2} \int \frac{\mathrm{d}^4 q}{(2\pi)^4} \frac{g_{\mu \nu} \left( \tau^{\mu \nu}_{VV} \left( q, v \right) + \tau_{VV, 0}^{\mu \nu} \left( q, v \right)\right)}{ q^2}. \label{eq:nucleon_EM}
\end{equation}
\end{linenomath}
The pure imaginary term $2 i \pi \delta \left( v \cdot r \right) \Delta_\mathrm{em} m_N$ cancels the nucleon pole singularity from ${t}_{AVV}$. The hadronic objects ${t}_{AVV}$ ($\tau_{VV}$) are defined in terms of the axial-vector quark bilinear and two vector currents (two vector currents) in Ref.~\citep{Cirigliano:2024nfi}.

We describe the evaluation of $\Delta\overline{g}_A^{\rm QCD + QED}$ in Appendix~\ref{app:AVV_nucleon_state3pt}, while the contribution from the nucleon two-current correlation function $2 \overline{m}{t}_{VP} - {\overline t}_{VA}$ is presented in Appendix~\ref{app:AVV_nucleon_state2pt}.

\subsection[\appendixname~\thesubsection]{Contribution from three-current correlation functions} \label{app:AVV_nucleon_state3pt}

The nucleon-state contribution from three-current correlation functions in Eq.~(\ref{eq:gA_extra_one_loop_AVV}) can be expressed as
\begin{linenomath}
\begin{align}
	&\frac{\alpha}{8 \pi} \int \frac{\mathrm{d} Q^2}{\beta m_N^2} \left[ \frac{1 - 2 \beta^3 + \beta^4}{1+ \beta} F_1^p F_1^n + \frac{1-\beta}{1+\beta} \left( 1 + \frac{\beta}{2} \right) \left[ F_1^p F_2^n + F_1^n F_2^p \right] + \frac{1 + \beta + \beta^3}{ \left( 1 + \beta \right)^2} F_2^p F_2^n \right] \nonumber \\
	&= -0.0127(2) \%,\label{eq:tVVA_form_factors}
\end{align}
\end{linenomath}
with the dimensionless parameter $\beta = \sqrt{1 + \frac{4 m_N^2}{Q^2}}$, exploiting the standard notations for nucleon form factors and fits to the experimental data from Ref.~\citep{Borah:2020gte}.

\subsection[\appendixname~\thesubsection]{Contribution from two-current correlation functions} \label{app:AVV_nucleon_state2pt}

We express the contribution from ${\overline t}_{VA}$ in terms of the nucleon form factors as
\begin{linenomath}
\begin{equation}
	- e^2 \int \frac{i \mathrm{d}^4 q}{(2\pi)^4} \frac{{\overline t}_{VA} \left(q, v \right)}{q^2} \to - \frac{\alpha}{12 \pi} \int \frac{\mathrm{d} Q^2}{m_N^2} \frac{1-\beta}{1+\beta} \left( \beta + \frac{1}{2} \right) \frac{G_M^V F_A}{g_A^{\left( 0 \right)}} = 0.144(3) \%, \label{eq:tVA_form_factors}
\end{equation}
\end{linenomath}
exploiting the nucleon axial-vector form factor from Refs.~\citep{MINERvA:2023avz,Tomalak:2026wsu}.

Corrections from the integral of $t_{VP}$ beyond the leading pion-pole contribution ${t}^{1\pi}_{VP}$ are chirally suppressed~\citep{Cirigliano:2024nfi}, i.e.,
\begin{linenomath}
\begin{equation}
	e^2 \int \frac{i \mathrm{d}^4 q}{(2\pi)^4} \frac{2 \overline{m} \left( {t}_{VP} \left(q, v \right) - {t}^{1\pi}_{VP} \left(q, v \right)\right)}{ q^2 - \lambda_\gamma^2} = \frac{\alpha}{\pi} \mathcal{O} \left( \frac{m^2_\pi}{m^2_N}, \frac{m^2_\pi}{\left( 4 \pi F_\pi \right)^2} \right) \lesssim 0.006\%. \label{eq:tVP_estimate}
\end{equation}
\end{linenomath}

It is instructive to analyze also the expression in terms of the nucleon form factors for the nucleon-state contribution from ${t}_{VP}$,
\begin{linenomath}
\begin{equation}
	e^2 \int \frac{i \mathrm{d}^4 q}{(2\pi)^4} \frac{2 \overline{m}{t}_{VP} \left(q, v \right)}{ q^2 - \lambda_\gamma^2} \to \frac{\alpha}{4\pi} \int \frac{\mathrm{d} Q^2}{Q^2} \left( \frac{1 - \beta}{1 + \beta} F_1^V + \frac{5 + 4 \beta}{3 \left( 1 + \beta \right)^2} F_2^V \right)\frac{F_A - \frac{Q^2}{2m_N^2} F_P}{g_A^{\left( 0 \right)}}. \label{eq:tVP_form_factors}
\end{equation}
\end{linenomath}
The $F_2^V$ term yields $0.046\%$. The leading pion-pole contribution ${t}^{1\pi}_{VP}$ is included as a part of the nucleon form factors. Subtracting it from Eq.~(\ref{eq:tVP_form_factors}), we obtain the finite nucleon-state contribution from ${t}_{VP}$: $0.078\%$.

\section[\appendixname~\thesection]{Alternative decomposition and Ward identities} \label{app:Ward_identity}

In this Appendix, we present the evaluation of the nucleon-state contribution $\delta g_A^{\rm extra, N}$ to $\delta g_A^{\rm extra}$ expressed in terms of alternative hadronic objects as
\begin{linenomath}
\begin{align}
	&\frac{\delta g_A^{\rm extra}}{g_A^{\left( 0 \right)}} =  e^2 \int \frac{i \mathrm{d}^4 q}{(2\pi)^4} \frac{2 \overline{m}{t}_{VP} \left(q, v \right)}{ q^2 - \lambda_\gamma^2} + \frac{\Delta_\mathrm{em} m_N}{m_N} - \frac{\alpha}{8 \pi}\nonumber \\
	&+ \left[ \frac{e^2 \overline{m} b_\rho b_\lambda}{g_A^{\left( 0 \right)}}  \frac{\partial }{\partial r_\lambda} \left( \hspace{-0.12cm} \int \hspace{-0.2cm} \frac{i \mathrm{d}^4 q}{(2\pi)^4} \frac{g_{\mu \nu} \left( {t}_{PVV}^{\mu \nu \rho} \left( r, q, v \right) + {t}_{PVV, 0}^{\mu \nu \rho} \left( r, q, v \right)\right)}{q^2 - \lambda^2_\gamma} \right) + 2 i \pi \delta \left( v \cdot r \right) \Delta_\mathrm{em} m_N \right]\Bigg |_{r_\lambda = 0},\label{eq:gA_extra_one_loop_PVV}
\end{align}
\end{linenomath}
with the momentum insertion in the pseudoscalar density $r_\lambda$. The hadronic objects ${t}_{PVV}$ are defined in terms of the pseudoscalar quark bilinear and two vector currents in Ref.~\citep{Cirigliano:2024nfi}.

We describe the evaluation of the integral from the nucleon three-current correlation functions in Appendix~\ref{app:PVV_nucleon_state3pt}, while the contribution from the nucleon two-current correlation functions is presented in Appendix~\ref{app:PVV_nucleon_state2pt}. In Appendix~\ref{app:summary}, we compare the results from Appendix~\ref{app:nucleon_state} and Appendix~\ref{app:Ward_identity}.

\subsection[\appendixname~\thesubsection]{Contribution from three-current correlation functions} \label{app:PVV_nucleon_state3pt}

The nucleon-state contribution from three-current correlation functions in Eq.~(\ref{eq:gA_extra_one_loop_PVV}) can be expressed as
\begin{linenomath}
\begin{align}
	&\frac{\alpha}{8 \pi} \int \frac{\mathrm{d} Q^2}{\beta m_N^2} \frac{1 - \beta}{1 + \beta} \left[  \left( 1 - \beta \right) \left( 1 + \beta \right)^2 F_1^p F_1^n + \left( 1 - \beta -  \frac{3}{2} \beta^2 \right) \left[ F_1^p F_2^n + F_1^n F_2^p \right] - \left( 1 + 2 \beta \right) F_2^p F_2^n \right] \nonumber \\
	&= -0.1558(5) \%.\label{eq:tVVP_form_factors}
\end{align}
\end{linenomath}

\subsection[\appendixname~\thesubsection]{Contribution from two-current correlation functions} \label{app:PVV_nucleon_state2pt}

Corrections from $t_{VP}$ are described in Appendix~\ref{app:AVV_nucleon_state2pt}, while additional two-current corrections correspond to the relative Born contribution to the isospin-averaged nucleon mass. We update the value according to Ref.~\citep{Tomalak:2018dho} $0.017(1)\%$ with a new result $0.0166(2)\%$ that accounts for recent progress in extracting the nucleon vector form factors~\citep{Borah:2020gte}.

\subsection[\appendixname~\thesubsection]{Discussion of Ward identities} \label{app:summary}

After adding the nucleon-state contribution of Eq.~(\ref{eq:tVVP_form_factors}) and the constant term in Eq.~(\ref{eq:gA_extra_one_loop_PVV}) to the nucleon-state terms in the isospin-averaged nucleon mass shift, the difference to Eq.~(\ref{eq:gA_extra_one_loop_AVV}) is $0.30\%$ that is an expected order of magnitude for inelastic contributions. We exploit this number for uncertainty estimates.

%%%%%%%%%%%%%%%%%%%%%%%%%%%%%%%%%%%%%%%%%%
\begin{adjustwidth}{-\extralength}{0cm}

\reftitle{References}

% ACS format
\isAPAandChicago{}{%

}

\PublishersNote{}
\end{adjustwidth}

\end{document}